\documentclass[final,5p,times,twocolumn,numbers]{elsarticle}

\usepackage[T1]{fontenc}
\usepackage[utf8]{inputenc}

\usepackage{amsmath}
\usepackage{amssymb}
\usepackage{bm}
\usepackage{xcolor}
\usepackage{soul}
\usepackage[normalem]{ulem}
\usepackage{cuted}
\usepackage{flushend}
\usepackage{multicol}
\usepackage{subcaption}
\usepackage{tikz}
\usetikzlibrary{decorations.pathmorphing}
\usepackage{float}
\usepackage{comment}

\usepackage{hyperref}
\hypersetup{
    colorlinks=true,
    linkcolor=red,
    filecolor=magenta,
    citecolor=blue,
    urlcolor=blue
}

\newcommand*{\emi}[1]{{\color{blue}#1}}

\journal{Physics Letters B}

\begin{document}


\begin{frontmatter}

\title{Scaling Solutions of Matter Form Factors in\\
Asymptotically Safe Quantum Gravity}

\author[first,second]{Alfio M. Bonanno}
\ead{alfio.bonanno@inaf.it}

\author[third,fourth]{Diego Buccio}
\ead{buccio@thphys.uni-heidelberg.de}

\author[first,second,fifth]{Emiliano M. Glaviano}
\ead{emiliano.glaviano@inaf.it}

\author[fourth]{Frank Saueressig}
\ead{f.saueressig@science.ru.nl}

\affiliation[first]{
organization={INAF Osservatorio Astrofisico di Catania},
addressline={Via S. Sofia 78},
city={Catania},
postcode={95123},
country={Italy}
}

\affiliation[second]{
organization={INFN, Sezione di Catania},
addressline={Via Santa Sofia 64},
city={Catania},
postcode={95123},
country={Italy}
}

\affiliation[third]{
organization={Institut f\"ur Theoretische Physik, Heidelberg University},
addressline={Philosophenweg 16},
city={Heidelberg},
postcode={69120},
country={Germany}
}

\affiliation[fourth]{
organization={Institute for Mathematics, Astrophysics and Particle Physics, Radboud University},
addressline={Heyendaalseweg 135},
city={Nijmegen},
postcode={6525 AJ},
country={The Netherlands}
}

\affiliation[fifth]{
organization={Dipartimento di Fisica e Astronomia, Universit\`a di Catania},
addressline={Via S. Sofia 64},
city={Catania},
postcode={95123},
country={Italy}
}

\begin{abstract}
We investigate the renormalization group flow of a gravity--matter system in which a scalar field is minimally coupled to Einstein gravity and its kinetic term is given by a scale-dependent form factor $f_\Lambda(-\Box)$. Employing the Wilsonian proper-time flow equation, we derive a closed integro-differential equation that encodes the dependence of the form factor on the UV cutoff $\Lambda$. We solve the resulting fixed-point problem with a pseudospectral discretization and find a non-trivial fixed point for which $f_\ast(-\Box)$ departs from the canonical $-\Box$ behavior. Linearizing the flow about this solution yields a discrete spectrum of perturbations and a corresponding set of critical exponents, indicating a non-trivial scaling structure in this non-local sector compatible with asymptotic safety. We also observe that the form factor becomes local once the UV cutoff is removed, suggesting that the bare action associated with this fixed point is local in the scalar two-point sector.
\end{abstract}


\end{frontmatter}



\section{Introduction}
\label{introduction}
It is commonly accepted that general relativity constitutes an effective field theory (EFT) that works below a certain UV-cutoff scale $\Lambda$ only. Quantum gravity research seeks to extend this EFT to a quantum theory that is valid on all length scales. Phrased differently, one is looking for ways in which the UV-cutoff can be sent to infinity without introducing unphysical divergences. Within the framework of quantum field theory, this leads to the central idea of the asymptotic safety program \cite{Weinberg:1980gg,Reuter:2019byg}: the high-energy completion of gravity is provided by a fixed point of the theory's renormalization group (RG) flow. If the flow approaches the fixed point, the limit $\Lambda \rightarrow \infty$ can be taken in a controlled and predictive way.

Traditionally, investigating the existence of suitable fixed points in gravity and gravity-matter systems is based on approximate solutions of the Wetterich equation \cite{Wetterich:1992yh,Morris:1993qb,Ellwanger:1993mw} adapted to gravity \cite{Reuter:1996cp}. This equation studies the RG flow and fixed points of the effective average action $\Gamma_k$ with respect to the IR-cutoff $k$ \cite{Dupuis:2020fhh}. On this basis, the existence of a suitable fixed point has been established for pure gravity, where the fixed point is called the Reuter fixed point, and also many phenomenologically interesting gravity-matter systems \cite{Reuter:2019byg,Percacci:2017fkn,Pereira:2019dbn,Bonanno:2020bil,Pawlowski:2020qer,Morris:2022btf,Eichhorn:2022gku,Saueressig:2023irs,DAngelo:2025yoy}, including (subsections of) the standard model of particle physics \cite{Reuter:2004nv,Shaposhnikov:2009pv,Eichhorn:2017ylw,Pastor-Gutierrez:2022nki,Eichhorn:2025sux}. In particular, gravity coupled to scalar matter has been investigated extensively \cite{Eichhorn:2012va,Dona:2013qba,Dona:2015tnf,Christiansen:2017gtg,Becker:2017tcx,Alkofer:2018fxj,Pawlowski:2018ixd,deBrito:2021pyi,Laporte:2021kyp,Knorr:2022ilz,Sen:2022xlp,deBrito:2023myf}, nurturing the expectation that a fixed point similar to the one appearing in pure gravity exists in this setting.

An essential ingredient in understanding the properties and implications of these fixed points are their form factors \cite{Knorr:2019atm,Knorr:2022dsx}. These provide a compact way of encoding the momentum dependence of infinitely many higher-derivative operators and, as such, constitute a natural language for quantum corrections in gravity and gravity-matter systems. In the asymptotic safety program they have become a pathway for going beyond finite-dimensional truncations, allowing one to track the full momentum dependence of propagators and vertices. Moreover, they are essential when addressing questions related to the resolution of spacetime singularities \cite{Buoninfante:2018xiw,Bosma:2019aiu}, connections between quantum gravity programs \cite{Knorr:2021iwv}, the construction of asymptotically safe amplitudes \cite{Draper:2020bop,Draper:2020knh,Chiesa:2026tlz,Knorr:2026vax}, and Lorentzian unitarity 
\cite{Becker:2017tcx,Platania:2020knd,Knorr:2021niv,Bonanno:2021squ,Fehre:2021eob,Knorr:2022dsx,Braun:2022mgx,Pawlowski:2023gym,Pawlowski:2025etp}.
At the same time, functional RG methods have increasingly been applied to genuinely non-local models, where the RG flow acts directly on momentum-dependent interactions rather than on a finite set of local couplings; see, e.g., recent FRG studies of non-local scalar theories \cite{Bonanno:2025enn} and references therein.

Motivated by these developments, we study a gravity--matter system in which the scalar kinetic term is promoted to a scale-dependent form factor. Our goal is to determine whether the coupled system admits non-trivial fixed points, compute the (non-local) form factor and to extract its scaling exponents. In variance with the bulk of the asymptotic safety literature, we formulate the RG flow in terms of the UV-scale dependent Wilsonian action $S_\Lambda$. The dependence of $S_\Lambda$ on the UV scale $\Lambda$ is obtained from a Wilsonian coarse-graining scheme, formulated using a proper-time (PT) regulator \cite{Bonanno:2004sy,Bonanno:2019ukb,Bonanno:2025tfj,Glaviano:2024hie}. From the perspective of asymptotic safety, this has the advantage that one can access the fixed point action $\lim_{\Lambda \rightarrow \infty} S_\Lambda \equiv S_*$. In this way one is able to draw conclusions whether the RG fixed point corresponds to a bare action which is local.\footnote{In principle, $S_*$ can also be obtained from the fixed functional of the effective average action $\Gamma_*$. This requires solving the reconstruction problem \cite{Manrique:2008zw,Manrique:2009tj,Morris:2015oca,Fraaije:2022uhg}. However, this is technically quite involved as it requires both finding the form factors in $\Gamma_*$ and subsequently constructing the explicit mapping $\Gamma_* \mapsto S_*$.} 

The scalar form factor constructed in this work is a key step towards answering this structural question. Our explicit computation shows that the form factor is genuinely non-local as long as the UV-cutoff is kept finite. Taking the limit $\Lambda \rightarrow \infty$ removes the non-local terms, indicating that the RG fixed point corresponds to a local bare action. This constitutes the main new insight of our work.

The remainder of the letter is organized as follows. Section~\ref{sec:theory} summarizes the PT flow equation used in this work and introduces the truncation. Section~\ref{sec:PS} describes the pseudospectral implementation and presents the fixed-point solution. We then analyze the linearized flow around the fixed point and determine the critical exponents. Section~\ref{sec:conclusions} contains our conclusions.

\section{The gravity matter system and the proper time flow equation}\label{sec:theory}
\subsection{Proper Time Flow Equation}
A convenient Wilsonian coarse-graining scheme can be formulated using a proper-time (PT) regulator. In this setting one introduces a UV-scale $\Lambda$ delimiting the fluctuations already integrated out and studies the flow of the corresponding Wilsonian action $S_\Lambda$. We employ the PT Wilsonian flow \cite{Bonanno:2019ukb}
\begin{equation}
\label{PTFE}
\partial_t S_t[\phi]
= \frac{1}{2}\int_{0}^{\infty}\frac{ds}{s}\, r\!\left(s, Z_t\right)\,
\mathrm{STr}\!\left[e^{-s\,S_t^{(2)}}\right]\,,
\end{equation}
Here $t \equiv \ln(\Lambda/\Lambda_0)$, $\Lambda_0$ is an arbitrary normalization scale, $S_t^{(2)}$ is the Hessian of $S_t$, $Z_t$ denotes the wave-function renormalization(s),
and $\mathrm{STr}$ is the supertrace over all fields (including ghosts) with the appropriate statistics. Eq.\ \eqref{PTFE} requires choosing the coarse-graining kernel $r(s, Z_t)$. We employ the spectrally adjusted C-type proper-time kernel introduced in \cite{Bonanno:2019ukb},
\begin{equation}
\label{cutoff}
r(s,Z_t)=2\frac{(s\,m\,\Lambda^2 Z_t)^m}{\Gamma(m)}\,e^{-s\,m\,\Lambda^2 Z_t}\,,
\end{equation}
where $m>0$ controls the shape of the cutoff family $r$ in the interpolation region. 

Unlike PT equations formulated for an IR-regulated 1PI effective action, where they arise as a one-loop improvement and are not exact in general, Eq.~\eqref{PTFE} is meant to define a Wilsonian RG scheme. In particular, it can be interpreted as an exact Wilsonian flow whenever the PT kernel corresponds to an underlying blocking transformation (equivalently, to a generator of infinitesimal field redefinitions along the flow) \cite{deAlwis:2017ysy,Bonanno:2019ukb}. We adopt this Wilsonian viewpoint throughout.

\subsection{The setting}
We consider a scalar field minimally coupled to Einstein gravity and study how quantum-gravitational fluctuations renormalize the scalar two-point function. To retain the full momentum dependence generated by the RG flow, we parameterize the quadratic part of the scalar action in terms of a running form factor $f_t(-\Box)$. Our truncation for the Wilsonian action reads
\begin{equation}
\label{action}
    S_t=\int d^4x \sqrt{g}\left[
    -\frac{1}{16\pi G_t} R
    +\frac12\,\phi\, Z_t^{(s)}\, f_t(-\Box)\,\phi\right] .
\end{equation}
Here $G_t$ denotes the scale-dependent Newton's coupling and $Z_t^{(s)}$ is the wave function renormalization of the scalar. Moreover, we fix the scalar field renormalization by imposing the normalization condition $f_t'(0)=1$, where a prime denotes a derivative with respect to $(-\Box)$. The inclusion of a cosmological constant term is straightforward; for simplicity, we set it to zero in the present work.

To compute the flow, we employ the linear split $g_{\mu\nu}=\bar g_{\mu\nu}+h_{\mu\nu}$ and choose the de Donder gauge fixing
\begin{equation}
S_t^{\rm gf}=\frac{1}{32\pi G_t}\int d^dx \sqrt{\bar g}\,
F_\mu\,\bar g^{\mu\nu}\,F_\nu,
\quad
F_\nu=\bar\nabla^\mu\left(h_{\mu\nu}-\frac{1}{2}\bar g_{\mu\nu}h\right).
\end{equation}
Here $h\equiv \bar g^{\mu\nu}h_{\mu\nu}$ and we use the bar to indicate covariant quantities constructed from the background metric. The corresponding Faddeev--Popov ghost action is taken at the classical level (no running ghost couplings),
\begin{equation}
S^{\rm gh}=\int d^dx \sqrt{\bar g}\,\bar c_\mu\,\mathcal{M}^{\mu}{}_{\nu}\,c^\nu,
\end{equation}
where
\begin{equation}
   \mathcal{M}^{\mu}{}_{\nu}= 
   \bar \nabla^\mu \bar g^{\rho\sigma}g_{\rho\nu}\nabla_\sigma
   - \bar \nabla^\rho \bar g^{\mu\sigma}g_{\sigma\nu}\nabla_\rho
   -\bar \nabla^\rho \bar g^{\mu\sigma}g_{\rho\nu}\nabla_\sigma.
\end{equation}

\subsection{Flow equations}
The proper-time Wilsonian flow generates coupled evolution equations for the vertices of $S_t$. We define $n$-point vertices by
\begin{equation}
    S_t^{(\psi_1\cdots\psi_n)}
    \equiv \left.\frac{\delta^n S_t}{\delta\psi_1\cdots\delta\psi_n}\right|_{\phi=0,\,h=0}\,,
\end{equation}
where $\psi_i\in\{\phi,h_{\mu\nu}\}$. Taking functional derivatives of the PT flow equation yields flow equations for these vertices. In the present work we focus on the flows of $S_t^{(0)}$ and the scalar two-point function $S_t^{(\phi\phi)}$ which encode the running of the background Newton's coupling $G_t$ and the form factor $f_t(p^2)$, respectively. The running of $G_t$ is extracted using a background sphere while the form factor is obtained on a flat background $\bar g_{\mu\nu}=\delta_{\mu\nu}$ so that all vertices admit a momentum-space representation. In practice we evaluate the functional derivatives acting on the heat-kernel operator using the Dyson expansion of the heat kernel \cite{Codello:2012kq}. For the proper-time kernel Eq. \eqref{cutoff}, the proper-time integral can be carried out analytically.
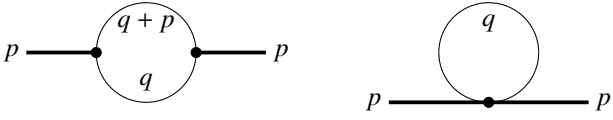
\begin{figure}[t]
		\centering
			\begin{tikzpicture}
				[>=stealth,scale=0.66,baseline=-0.1cm]
				\draw (-1,0) arc (180:0:1cm);
				\draw (-1,0) arc (180:360:1cm);
				\draw [line width=0.5mm] (-1,0) node[anchor=west] {} -- (-2.4,0);
				\draw [line width=0.5mm ] (2.4,0) -- (1,0) node[anchor=east] {};
				\draw[fill] (1,0) circle [radius=0.1];
				\draw[fill] (-1,0) circle [radius=1mm];
				\node [] at (-2.7,0) {$p$};
				\node [] at (2.7,0) {$p$};
				\node [] at (0,0.6) {$q+p$};
				\node [] at (0,-0.6) {$q$};
			\end{tikzpicture}
			\qquad
			\begin{tikzpicture}
				[>=stealth,scale=0.66,baseline=-0.1cm]
				\draw (-1,0) arc (180:0:1cm);
				\draw (-1,0) arc (180:360:1cm);
				\draw [line width=0.5mm](2,-1) node[anchor=west] {} -- (0,-1);
				\draw [line width=0.5mm](0,-1) -- (-2,-1) node[anchor=east] {};
				\draw[fill] (0,-1) circle [radius=1mm];
				\node [] at (-2.3,-1) {$p$};
				\node [] at (2.3,-1) {$p$};
				\node [] at (0,0.6) {$q$};
			\end{tikzpicture}
			\caption{Diagrams contributing to the scalar 2-point function:
				bubble (left) and tadpole (right).
				The thin lines correspond to fluctuation fields, the thick line stands for the external scalars with momentum $p$.
			}
			\label{fig:feynman}
	\end{figure}

The running of $G_t$ and $Z_t^{(s)}$ is conveniently encoded in the anomalous dimensions
\begin{equation}
\label{eqeta1}
\eta_t^{(g)}=\frac{\partial_t G_t}{G_t}\,,
\qquad
\eta_t^{(s)}=-\frac{\partial_t Z_t^{(s)}}{Z_t^{(s)}}\, .
\end{equation}

Introducing the dimensionless momentum variable
$x\equiv p^2/\Lambda^2$ and the dimensionless Newton coupling $g_t \equiv \Lambda^{d-2}G_t$, we define the dimensionless scalar wave-function factor
\begin{equation}
z_t^{(s)} \equiv \Lambda^{\eta_t^{(s)}} Z_t^{(s)}\,,
\end{equation}
and the dimensionless form factor through
\begin{equation}
\bar F_t(x)\equiv \Lambda^{-2+\eta_t^{(s)}} \, z_t^{\left(s\right)} \, f_t\left(p^2=x\Lambda^2\right)\equiv z_t^{(s)}\,F_t(x)\,.
\end{equation}
This implies
\begin{equation}
\label{def:dimlessff}
F_t(x)\equiv \Lambda^{-2} \, f_t(p^2=x\Lambda^2) \, . 
\end{equation}
With these conventions, the flow equations take the form
\begin{equation}\begin{split}
\label{ADeqF}
&\partial_t g_t=g_t\left(d-2+\eta_t^{(g)}\right)\,,\\
&\partial_t F_t(x)=\left(-2+\eta_t^{(s)}\right)F_t(x)+2x\,F_t'(x)+H(x,g_t)\,.
\end{split}\end{equation}
The background graviton anomalous dimension is given by
\begin{multline}
\eta_t^{(g)}=\frac{g_k}{3\left(4\pi\right)^{\frac{d}{2}-1}}\Bigg[-m^{\frac{d}{2}-1}\left(5d^2-3d+24\right)\frac{\Gamma{\left(m-\frac{d}{2}+1\right)}}{\Gamma{\left(m\right)}}+
\\
+\frac{2}{\Gamma{\left(\frac{d}{2}-1\right)}}\int_{0}^{\infty}{\frac{x^{\frac{d}{2}-2}}{\left(1+\frac{F_k{\left(x\right)}}{m}\right)^m}dx}\Bigg]\ ,
\end{multline}
while the loop contribution $H$ entering the flow of the dimensionless form factor reads
\begin{equation}
H\left(x,g_t\right)=g_t\frac{4\pi^\frac{1}{2}m^{1+m}}{\left(4\pi\right)^\frac{d}{2}d\Gamma\left(\frac{d-1}{2}\right)}\left(\kappa_1+\kappa_2\right),
\end{equation}
with
\begin{equation}\begin{split}
\label{flowHparts}
\kappa_1
&=\int_{0}^{\infty}\!dy\int_{-1}^{1}\!dQ\,
\frac{\left(1-Q^2\right)^{\frac{d-3}{2}}\,y^{\frac{d}{2}}\,A_k(x,Q,y)}
{\left(y^{\frac{3}{2}}+2Q\sqrt{x}\,y\right)^2\left(m+y\right)^{m+1}}\,,
\\
\kappa_2
&=\int_{0}^{\infty}\!dy\int_{-1}^{1}\!dQ\,
\frac{\left(1-Q^2\right)^{\frac{d-3}{2}}\,y^{\frac{d}{2}}\,B_k(x,Q,y)\,C_k(x,Q,y)}
{\left(y^{\frac{3}{2}}+2Q\sqrt{x}\,y\right)^2\left(m+y\right)^{m+1}}\,.
\end{split}\end{equation}
The two terms in $H$ correspond to the tadpole and self-energy contributions shown in Fig.~\ref{fig:feynman}. Here $y=q^2/\Lambda^2$ denotes the dimensionless loop momentum and $Q=\cos\theta$ parameterizes the relative angle between the external momentum $p$ and the loop momentum $q$. The explicit expressions for the functions $A_k(Q,y)$, $B_k(Q,y)$ and $C_k(Q,y)$ are lengthy and will be given in \ref{AppA}. They are obtained by evaluating the PT flow for the scalar two-point function using the Dyson expansion of the heat kernel \cite{Codello:2012kq} and the form-factor vertices, in close analogy to the construction employed in \cite{Knorr:2019atm}.

The scalar anomalous dimension $\eta_t^{(s)}$ follows from differentiating Eq.~\eqref{ADeqF} with respect to $x$,
evaluating at $x=0$ and using the normalization condition $F_t'(0)=1$:
\begin{equation}
\eta_t^{(s)}=-\,\left.\partial_x H(x,g_t)\right|_{x=0}\equiv -H^{(1,0)}(0,g_t)\,.
\end{equation}
The resulting closed expression is too lengthy to be displayed here.

\section{The non-trivial fixed-point solution}
\label{sec:PS}

\subsection{Form factor from the pseudospectral method}
The position of the non-trivial fixed point $g_*$ of the dimensionless Newton coupling $g_t$ follows from solving $\partial_t g_t=0$. For $d=4$, it can be written as
\begin{equation}
\label{eq:gstar}
g_\ast=\frac{12(m-1)\pi}{46m-(m-1)\int_{0}^{+\infty}\!\left(1+\frac{F_\ast(x)}{m}\right)^{-m}dx}\,.
\end{equation}

The fixed-point equation for the dimensionless form factor is obtained by setting $\partial_t F_t(x)=0$ in Eq.~\eqref{ADeqF}. Inserting Eq.~\eqref{eq:gstar} into the fixed-point condition for $F_\ast(x)$ yields a single integro-differential equation for the form factor. We solve this equation using a pseudospectral collocation method, following \cite{Knorr:2019atm}. To apply the method one needs the large-$x$ behavior of the solution. At a fixed point the scalar propagator scales as $G_\ast(p^2)\sim p^{-2+\eta_\ast^{(s)}}$, so $F_\ast(x)=1/G_\ast\sim x^\alpha$ with $\alpha\ge 1$ for $\eta_\ast^{(s)}<0$. By inserting a power-law ansatz into the fixed-point equation and expanding for $x\to\infty$ we find $\alpha\simeq 1.16$.

Following \cite{Knorr:2019atm}, we factor out the leading growth and compactify $x\in[0,\infty)$ to $[-1,1]$:
\begin{equation}
\label{eq:compact}
F_\ast(x)=(1+x)^\gamma\,\widetilde F_\ast\!\left(\frac{x-L}{x+L}\right)\,,
\end{equation}
where $\gamma>\alpha$ ensures that $\widetilde F_\ast$ is bounded. In practice we choose $\gamma=2$ to obtain rapidly convergent numerical quadratures and set $L=1$. The bounded function $\widetilde F_\ast$ is then expanded in Chebyshev polynomials of the first kind $T_n(x)$,
\begin{equation}
\label{eq:cheb}
\widetilde F_\ast\!\left(\frac{x-L}{x+L}\right)=\sum_{n=0}^{N} c_n\,
T_n\!\left(\frac{x-L}{x+L}\right)\,,
\end{equation}
and the fixed-point equation is enforced on a Gauss--Lobatto collocation grid \cite{article}. This yields a finite non-linear system for the coefficients $\{c_n\}$ together with $g_\ast$ and $\eta_\ast^{(s)}$, supplemented by the normalization condition $F_\ast'(0)=1$.

We study convergence by increasing the truncation order $N$ starting from $N=8$. To test the convergence of the results, we use the absolute relative difference $\delta F_\ast^{PS}\left(N,x\right)=\left|1-\frac{F_\ast\left(N,x\right)}{F_\ast\left(N+1,x\right)}\right|$ evaluated as a function of $N$ for different values of $x$. At $N=25$ the relative difference $\delta F_\ast^{PS}\left(N,x\right)\sim{10}^{-6}$ for all tested values of $x$, consequently the convergence of the pseudospectral approximation is reached. The same happens for the convergence of the values of $g_\ast$ and $\eta_\ast^{(s)}$ as a function of the truncation $N$. The relative difference at $N=25$ is around ${10}^{-6}$ for $g_\ast$ and ${10}^{-4}$ for $\eta_\ast^{\left(s\right)}$. Table~\ref{tablevalFP} summarizes the fixed-point data for representative values of $m$. 

\begin{table}[t]
\caption{Fixed-point values obtained at truncation order $N=25$ for different values of the cutoff parameter $m$.
We also list the large-$x$ scaling exponent $\alpha$ and the second relevant critical exponent $\theta_2$
reported below.}
\label{tablevalFP}
$\begin{array}{|c|cccc|}
\hline
 m & g_* & \eta _* & \alpha  & \theta_2 \\
\hline
 3 & 0.558 & -0.529 & 1.160 & 1.1950\\
 5 & 0.670 & -0.530 & 1.160 & 1.1970 \\
 10 & 0.754 & -0.530 & 1.160 & 1.1970 \\
\hline
\end{array}$
\end{table}

\begin{figure*}[t]
    \centering
    \begin{subfigure}[t]{0.45\textwidth}
        \centering
        \includegraphics[width=\textwidth]{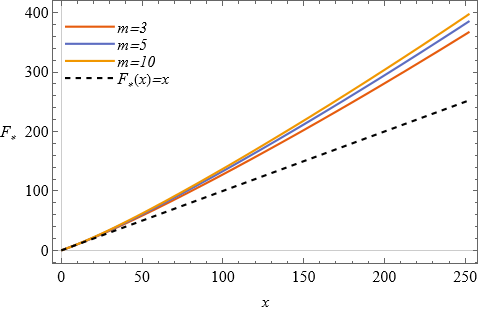}
        \caption{}
        \label{plotFshort}
    \end{subfigure}
    \hspace{1.8em}
    \begin{subfigure}[t]{0.5\textwidth}
        \centering
        \includegraphics[width=\textwidth]{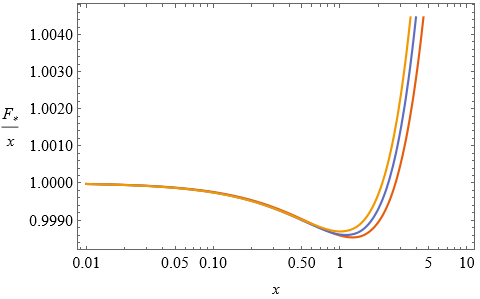}
        \caption{}
        \label{plotlogFshort}
    \end{subfigure}
    \caption{Fixed-point form factor at $N=25$ for different values of $m$ (left panel) and log--log plot of
    $F_\ast(x)/x$ (right panel). The solution follows $F_\ast(x)\simeq x$ up to $x\sim 1$ where there is a crossover. For large value $x$ the solutions follow a power law $F_*(x) \sim x^\alpha$ with $\alpha \simeq 1.16$. 
    \label{Fig2}}
\end{figure*}

The key properties of the non-trivial fixed point solution are summarized in Table \ref{tablevalFP} and Figure \ref{Fig2}. Figure~\ref{plotFshort} shows the fixed-point form factor at $N=25$ for different values of $m$ together with the reference behavior $F_\ast(x)=x$. The solution follows $F_\ast(x)\simeq x$ up to $x\sim 1$ and deviates at larger $x$, as also seen from the log--log plot of $F_\ast(x)/x$ in Fig.~\ref{plotlogFshort}. From the logarithmic derivative $\alpha(x)=xF_\ast'(x)/F_\ast(x)$ we extract $\alpha(x\to\infty)\simeq 1.16$. Moreover, $F_\ast(0)=0$, so the fixed-point mass gap $\mu_\ast^2=\left(Z_\ast^{(s)}\right)^{-1} f_\ast(p^2)\big|_{p=0}$ vanishes.

The fixed-point solution is naturally a function of the dimensionless variable $x=p^2/\Lambda^2$. In terms of this variable, the fixed-point form factor is genuinely non-trivial,
\begin{equation}
F_\ast(x)\neq x .
\end{equation}
In particular, at large dimensionless momentum we find
\begin{equation}\label{ffnontrivial}
F_\ast(x)\sim x^\alpha,
\qquad
\alpha\simeq 1.16 .
\end{equation}
Thus, in the fixed-point scaling regime, the scalar two-point kernel does not scale canonically. Schematically, one has
\begin{equation}
S_{\ast}^{(\phi\phi)}(p^2)
\sim (p^2)^\alpha ,
\end{equation}
up to the appropriate powers of the cutoff and wave-function normalization. Equivalently, the fixed-point propagator scales as
\begin{equation}
G_\ast(p^2)\sim (p^2)^{-\alpha},
\end{equation}
rather than as the canonical $1/p^2$. Since $\alpha$ is not an integer, this behavior may be interpreted as a mild non-locality of the fixed-point scaling kernel.


\subsection{Critical exponents}
To analyze the stability of the fixed point, we consider linear perturbations
\begin{equation}
\label{pertFP}
g_t=g_\ast+\delta g_t\,,\qquad
F_t(x)=F_\ast(x)+\delta F_t(x)\,.
\end{equation}
In \cite{Knorr:2019atm} it was argued that, in the absence of a cosmological constant, the only physically relevant deformation of $F_*$ is associated with the mass gap. This conclusion relies on parameterizing $F_t(x)=Z_t(x)\,(x+\mu_t^2/\Lambda^2)$ and absorbing $Z_t(x)$ by a field redefinition. Since we are interested in the full momentum dependence at the fixed point, we do not impose this restriction a priori and allow for general perturbations $\delta F_t(x)$.

We expand $\delta F_t$ in the same Chebyshev basis as the fixed-point solution,
\begin{equation}
\delta F_t(x)=(1+x)^\gamma \sum_{n=0}^N \delta c_n(t)\,
T_n\!\left(\frac{x-L}{x+L}\right)\,,
\end{equation}
insert Eq.~\eqref{pertFP} into the flow equations \eqref{ADeqF}, linearize about the fixed point, and evaluate on the collocation grid. This yields the finite-dimensional linear system
\begin{equation}
\partial_t\,\delta\mathbf{u}(t)=\mathbf{M}\,\delta\mathbf{u}(t)\,,
\qquad
\delta\mathbf{u}=(\delta g_t,\delta c_0,\ldots,\delta c_N)^{\rm T}\,,
\end{equation}
where $\mathbf{M}$ is the stability matrix. Its eigenvalues $\lambda_i$ determine the critical exponents via $\theta_i=-\lambda_i$. The normalization condition $F_t'(0)=1$ removes the redundant rescaling direction; at the linear level this implies $\delta F_t'(0)=0$.

For fixed $m$ we study the stability matrix for each truncation from $N=8$ to $N=25$. We find two relevant directions. These correspond to the background anomalous dimension $\theta_1$ and the canonical mass deformation  $\theta_2$. 
The second relevant exponent is well described by
\begin{equation}
\theta_2=2-\eta_\ast^{(s)}+X\,,
\end{equation}
where $X$ arises from the tadpole contribution in the linearized flow. For $m=3$ we have $\eta_\ast^{(s)}\simeq -0.529$ and $\theta_2\simeq 1.195$, implying
\begin{equation}
X \simeq \theta_2-(2-\eta_\ast^{(s)}) \simeq 1.195-2.529 \simeq -1.334\,.
\end{equation}
This is consistent with the value extracted directly from the linearized operator. All remaining critical exponents have $\mathrm{Re}\,\theta<0$, with the spectrum of $\mathbf{M}$ containing both real and complex conjugate pairs of eigenvalues. 

\begin{figure}[t]
        \centering
        \includegraphics[width=0.48\textwidth]{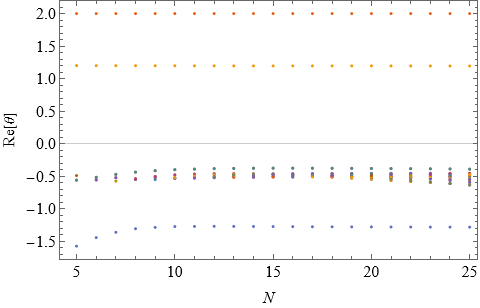}
        \caption{Plot of the critical exponents as function of the truncation $N$. The two relevant directions where Re $\theta > 0$ correspond to the orange and yellow dots. The irrelevant critical exponents are shown as dots with  Re $\theta < 0$. }
        \label{plotthetashort}%
\end{figure}

Increasing the truncation stabilizes the values of the two positive critical exponents and with $N=25$ the convergence is reached where the relative difference is around $10^{-6}$. Fig. \ref{plotthetashort} shows the positive critical exponents as functions of the truncation. At $N=25$ the resulting values of $\theta_2$ for several $m$ are listed in Table~\ref{tablevalFP} and are close to those found in \cite{Knorr:2019atm}.

For the negative critical exponents, when the truncation is increased, a new value adds to the previous determined set. This indicates that there are an infinite number of negative critical exponents. However, how to order the set of negative critical exponents is not uniquely determined, different criteria can be used (for example by ordering the real part, looking at the superposition of the respective eigenvectors, the module and so on) and the convergence cannot be studied easily. Fig. \ref{plotthetashort} shows the real part of some critical exponents with $m=3$ for different values of $N$. 

\subsection{Emergent locality}
A key feature of the fixed-point form factor $F_*(-\Box)$ is that it is non-trivial as long as the UV-cutoff $\Lambda$ is kept finite, see Eq.\ \eqref{ffnontrivial}. The UV-completion of an RG trajectory through this fixed point corresponds to taking the limit $\Lambda \rightarrow \infty$, keeping the physical momentum $p$ fixed. Thus, it is the limit
\begin{equation}
x=\frac{p^2}{\Lambda^2}\to 0 \, ,
\end{equation}
which is relevant for the reconstruction of the dimensionful Wilsonian form factor appearing in the bare action. Thus only the small-$x$ expansion of the fixed-point form factor is probed. Using the normalization condition $F_\ast'(0)=1$ and the analyticity of the solution at the origin, one may write
\begin{equation}
\label{Fexpansion}
F_\ast(x)
=
x\left[
1+\sum_{n=1}^{\infty}a_n x^n
\right].
\end{equation}
The series coefficients $a_n$ can be computed in terms of the $c_n$. Their precise value is unimportant for the argument though and thus we do not give their values.

Combining the expansion \eqref{Fexpansion} with the relation \eqref{def:dimlessff} this entails
\begin{equation}
f_*(p^2)=p^2\left[1+\sum_{n=1}^{\infty}a_n\left(\frac{p^2}{\Lambda^2}\right)^n\right].
\label{eq.locality1}
\end{equation}
Taking the limit $\Lambda \rightarrow \infty$ the infinite series in \eqref{eq.locality1} vanishes and one is left with an expression that is local and obeys the canonical scaling behavior\footnote{Note that this is not related to the concept of ''momentum locality'' \cite{Christiansen:2015rva}, which states that the RG flow of an $n$-point function at a given external momentum is dominated by loop momenta of the same order.}
\begin{equation}
    \label{eq.locality2}
     \lim_{\Lambda \rightarrow \infty} f_*(p^2) = p^2 \, . 
\end{equation}
This leads us to the following, profound observation: as long as the UV-cutoff $\Lambda$ is left finite the solution of the fixed point equation is non-local. If the UV-cutoff is sent to infinity, the non-locality is removed. The latter is the limit which corresponds to the UV-completion of an RG trajectory approaching the fixed point. On this basis, we conclude that the bare action associated with our fixed point solution is local. We stress that this property is independent of the precise details of the form factor. It merely hinges on the analyticity of $F_*(x)$ at the origin.

\section{Summary and conclusions}
\label{sec:conclusions}

In this work we studied a scalar field minimally coupled to Einstein gravity, allowing for a general momentum-dependent kinetic term encoded in a running form factor $f_\Lambda(-\Box)$. The renormalization-group flow was formulated for the UV-regulated Wilsonian action $S_\Lambda$ using a proper-time flow equation with a spectrally adjusted kernel \cite{Bonanno:2019ukb}. Building on the Dyson expansion techniques of \cite{Codello:2012kq}, we implemented a vertex expansion of the proper-time flow and derived coupled flow equations for the $n$-point functions $S_\Lambda^{(n)}$. In particular, the flow of the scalar two-point kernel (and thus of the form factor) is an integro-differential equation with two contributions, corresponding to the self-energy and tadpole diagrams.

We solved the fixed-point equation and the linearized flow around it with a pseudospectral method, determining the form factor for all values of the dimensionless momentum $x=p^2/\Lambda^2$. The resulting fixed point for the dimensionless Newton coupling is a proper-time analogue of the Reuter fixed point. The deviation from the classical behavior $F_\ast(x)=x$ is mild at intermediate momenta and becomes visible only at large $x$, where the solution crosses over to a power law $F_\ast(x)\sim x^{\alpha}$ with $\alpha\simeq 1.16$ for the regulator choices investigated. At the fixed point we find a negative scalar anomalous dimension, $\eta_\ast^{(s)}\simeq -0.53$ (e.g. for $m=10$), and $F_\ast(0)=0$, implying a vanishing fixed-point mass parameter in the present truncation. 

It is instructive to compare these findings with the results \cite{Knorr:2019atm} that compute the same form factor using the Wetterich equation for the effective average action. In that case, it was found that $\eta_\ast^{(s)} = -0.176$ and $\alpha\simeq 0.94$. Thus, we find the same signs for $\eta_\ast^{(s)}$ and $\alpha$, but a larger magnitude in both cases.

Linearizing the flow about the fixed point we obtained the spectrum of critical exponents. Besides the background anomalous dimension $\theta_1=2$, we find one additional relevant exponent associated with the canonical mass deformation, $\theta_2\simeq 1.195$. The remaining exponents have negative real parts and include complex conjugate pairs. The number of relevant directions is stable under increasing the pseudospectral truncation order. We repeated the same computation using a type B regulator \cite{Bonanno:2019ukb}, and the results are qualitatively consistent. Interpreting the form factor as encoding an infinite set of higher-derivative operators, these results provide evidence that the coupled gravity--scalar system admits a non-trivial fixed point with a finite number of UV-attractive deformations in this momentum-dependent sector.

We also find that the scalar form factor computed in this work reduces to a local expression once the UV-cutoff is sent to infinity. This feature solely hinges on the analytic properties of the form factor at the origin and is thus largely independent of its precise structure. We interpret this as first-hand evidence that the bare action corresponding to the Reuter fixed point at the level of the path integral is local.

\section*{Acknowledgements}
\noindent We thank Marc Schiffer and Omar Zanusso for useful discussions. D.B. and E.G. thank IMAPP for the hospitality. D.B. acknowledges Fondazione Angelo Della Riccia for the financial support. D.B. also acknowledges the European Research Council’s (ERC) support under the European Union’s Horizon 2020 research and innovation program Grant agreement No. 101170215 (ProbeQG).

\appendix

\begin{strip}
\section{Complete flow equations}
 \label{AppA}
The expressions for $A_k$, $B_k$, and $C_k$ in Eq. (\ref{flowHparts}) are given by
 
\begin{align}
\begin{split}
&A_k\left(x,Q,y\right)=\left\{y^\frac{3}{2}\left(4Q\sqrt x+\sqrt y\right)d\left(d^2+2d-4\right)+xy\left[\left(4d^3+6d^2-4d-16\right)Q^2+4d\right]+8\left(3d-4\right)Qx^\frac{3}{2}\sqrt y+8\left(d-2\right)x^2\right\}F_k\left(x\right)+\\
&+x\left\{2y\left[\left(d^2-6d+8\right)Q^2-2d\right]+8\sqrt{xy}\left(4-3d\right)Q-8\left(d-2\right)x\right\}F_k{\left(2Q\sqrt{xy}+x+y\right)}+\\
&+x\sqrt y\Bigg\{-y^\frac{3}{2}\left[2\left(d^2-6d+8\right)Q^2+4\left(d^2+d-4\right)\right]+4x^\frac{1}{2}yQ\left[4\left(2-d^2\right)-\left(d^2-6d+8\right)Q^2\right]+8x\sqrt y\left[\left(d-2\right)-2\left(d-1\right)dQ^2\right]+\\
&+16\left(d-2\right)x^\frac{3}{2}Q\Bigg\}F_k^\prime\left(x\right)
\end{split}
\\
&B_k\left(x,Q,y\right)=-\frac{2m}{\left(1+m\right)\left(y-F_k\left(x+2Q\sqrt{xy}+y\right)\right)}\left[1-\left(\frac{m+y}{m+F_k\left(x+2Q\sqrt{xy}+y\right)}\right)^{1+m}\right]\\
\begin{split}
&C_k\left(x,Q,y\right)=\left\{-d^2y^2-2d\left(2+d\right)Q\sqrt x y^\frac{3}{2}-\left[2d+\left(8+d\left(2+d\right)\right)Q^2\right]xy+4\left(d-4\right)Qx^\frac{3}{2}\sqrt y+4\left(d-2\right)x^2\right\}F_k^2\left(x\right)+\\
&+\left(2-d\right)\sqrt{xy}\left(2dQy+2\left(4+d\right)Q^2\sqrt{xy}+16Qx+8x^2\right)F_k\left(x\right)F_k\left(x+2Q\sqrt x\sqrt y+y\right)+\\
&+x\left\{\left[2d-\left(d-4\right)\left(d-2\right)Q^2\right]y+4\left(3d-4\right)Q\sqrt{xy}+4\left(d-2\right)x\right\}F_k^2\left(x+2Q\sqrt x\sqrt y+y\right)
\end{split}
\end{align}
\end{strip}


\bibliographystyle{elsarticle-num} 
\biboptions{sort&compress}
\bibliography{refFormFact.bib}

@article{Weinberg:1980gg,
 author               = {Weinberg, Steven},
 journal              = {General Relativity: An Einstein centenary survey, Eds. Hawking, S.W., Israel, W; Cambridge University Press},
 pages                = {790--831},
 slaccitation         = {%\%CITATION = INSPIRE-159043;\%\%},
 title                = {{Ultraviolet divergences in quantum theories of gravitation}},
 year                 = {1979},
 }

@book{Reuter:2019byg,
    author = "Reuter, Martin and Saueressig, Frank",
    title = "{Quantum Gravity and the Functional Renormalization Group}: {The Road towards Asymptotic Safety}",
    isbn = "978-1-107-10732-8, 978-1-108-67074-6",
    publisher = "Cambridge University Press",
    month = "1",
    year = "2019"
}

@article{Eichhorn:2022gku,
    author = "Eichhorn, Astrid and Schiffer, Marc",
    title = "{Asymptotic safety of gravity with matter}",
    eprint = "2212.07456",
    archivePrefix = "arXiv",
    primaryClass = "hep-th",
    month = "12",
    year = "2022"
}

@article{Laporte:2021kyp,
    author = "Laporte, Cristobal and Pereira, Antonio D. and Saueressig, Frank and Wang, Jian",
    title = "{Scalar-tensor theories within Asymptotic Safety}",
    eprint = "2110.09566",
    archivePrefix = "arXiv",
    primaryClass = "hep-th",
    doi = "10.1007/JHEP12(2021)001",
    journal = "JHEP",
    volume = "12",
    pages = "001",
    year = "2021"
}

@article{Alkofer:2018fxj,
    author = "Alkofer, Nat{\'a}lia and Saueressig, Frank",
    title = "{Asymptotically safe $f(R)$-gravity coupled to matter I: the polynomial case}",
    eprint = "1802.00498",
    archivePrefix = "arXiv",
    primaryClass = "hep-th",
    doi = "10.1016/j.aop.2018.07.017",
    journal = "Annals Phys.",
    volume = "396",
    pages = "173--201",
    year = "2018"
}

@article{Pawlowski:2018ixd,
    author = "Pawlowski, Jan M. and Reichert, Manuel and Wetterich, Christof and Yamada, Masatoshi",
    title = "{Higgs scalar potential in asymptotically safe quantum gravity}",
    eprint = "1811.11706",
    archivePrefix = "arXiv",
    primaryClass = "hep-th",
    doi = "10.1103/PhysRevD.99.086010",
    journal = "Phys. Rev. D",
    volume = "99",
    number = "8",
    pages = "086010",
    year = "2019"
}

@article{Pastor-Gutierrez:2022nki,
    author = "Pastor-Guti{\'e}rrez, {\'A}lvaro and Pawlowski, Jan M. and Reichert, Manuel",
    title = "{The Asymptotically Safe Standard Model: From quantum gravity to dynamical chiral symmetry breaking}",
    eprint = "2207.09817",
    archivePrefix = "arXiv",
    primaryClass = "hep-th",
    doi = "10.21468/SciPostPhys.15.3.105",
    journal = "SciPost Phys.",
    volume = "15",
    number = "3",
    pages = "105",
    year = "2023"
}

@article{Draper:2020bop,
    author = "Draper, Tom and Knorr, Benjamin and Ripken, Chris and Saueressig, Frank",
    title = "{Finite Quantum Gravity Amplitudes: No Strings Attached}",
    eprint = "2007.00733",
    archivePrefix = "arXiv",
    primaryClass = "hep-th",
    doi = "10.1103/PhysRevLett.125.181301",
    journal = "Phys. Rev. Lett.",
    volume = "125",
    number = "18",
    pages = "181301",
    year = "2020"
}

@article{Draper:2020knh,
    author = "Draper, Tom and Knorr, Benjamin and Ripken, Chris and Saueressig, Frank",
    title = "{Graviton-Mediated Scattering Amplitudes from the Quantum Effective Action}",
    eprint = "2007.04396",
    archivePrefix = "arXiv",
    primaryClass = "hep-th",
    doi = "10.1007/JHEP11(2020)136",
    journal = "JHEP",
    volume = "11",
    pages = "136",
    year = "2020"
}

@article{Buoninfante:2018xiw,
    author = "Buoninfante, Luca and Koshelev, Alexey S. and Lambiase, Gaetano and Mazumdar, Anupam",
    title = "{Classical properties of non-local, ghost- and singularity-free gravity}",
    eprint = "1802.00399",
    archivePrefix = "arXiv",
    primaryClass = "gr-qc",
    doi = "10.1088/1475-7516/2018/09/034",
    journal = "JCAP",
    volume = "09",
    pages = "034",
    year = "2018"
}

@article{Bosma:2019aiu,
    author = "Bosma, Lando and Knorr, Benjamin and Saueressig, Frank",
    title = "{Resolving Spacetime Singularities within Asymptotic Safety}",
    eprint = "1904.04845",
    archivePrefix = "arXiv",
    primaryClass = "hep-th",
    doi = "10.1103/PhysRevLett.123.101301",
    journal = "Phys. Rev. Lett.",
    volume = "123",
    number = "10",
    pages = "101301",
    year = "2019"
}

@article{Knorr:2021iwv,
    author = "Knorr, Benjamin and Ripken, Chris and Saueressig, Frank",
    title = "{Form Factors in Quantum Gravity: Contrasting non-local, ghost-free gravity and Asymptotic Safety}",
    eprint = "2111.12365",
    archivePrefix = "arXiv",
    primaryClass = "hep-th",
    doi = "10.1393/ncc/i2022-22028-5",
    journal = "Nuovo Cim. C",
    volume = "45",
    number = "2",
    pages = "28",
    year = "2022"
}

@article{Platania:2020knd,
    author = "Platania, Alessia and Wetterich, Christof",
    title = "{Non-perturbative unitarity and fictitious ghosts in quantum gravity}",
    eprint = "2009.06637",
    archivePrefix = "arXiv",
    primaryClass = "hep-th",
    doi = "10.1016/j.physletb.2020.135911",
    journal = "Phys. Lett. B",
    volume = "811",
    pages = "135911",
    year = "2020"
}

@article{Bonanno:2025enn,
    author = "Bonanno, Alfio M. and Haridev, S. R. and Narain, Gaurav",
    title = "{RG studies of scalar-field models of long-range interactions}",
    eprint = "2511.22666",
    archivePrefix = "arXiv",
    primaryClass = "hep-th",
    doi = "10.1103/tfqd-hvmy",
    journal = "Phys. Rev. D",
    volume = "113",
    number = "4",
    pages = "045021",
    year = "2026"
}

@article{Wetterich:1992yh,
    author = "Wetterich, Christof",
    title = "{Exact evolution equation for the effective potential}",
    eprint = "1710.05815",
    archivePrefix = "arXiv",
    primaryClass = "hep-th",
    reportNumber = "HD-THEP-92-61",
    doi = "10.1016/0370-2693(93)90726-X",
    journal = "Phys. Lett. B",
    volume = "301",
    pages = "90--94",
    year = "1993"
}

@article{Morris:1993qb,
    author = "Morris, Tim R.",
    title = "{The Exact renormalization group and approximate solutions}",
    eprint = "hep-ph/9308265",
    archivePrefix = "arXiv",
    reportNumber = "CERN-TH-6977-93, SHEP-92-93-27",
    doi = "10.1142/S0217751X94000972",
    journal = "Int. J. Mod. Phys. A",
    volume = "9",
    pages = "2411--2450",
    year = "1994"
}

@article{Manrique:2008zw,
    author = "Manrique, Elisa and Reuter, Martin",
    title = "{Bare Action and Regularized Functional Integral of Asymptotically Safe Quantum Gravity}",
    eprint = "0811.3888",
    archivePrefix = "arXiv",
    primaryClass = "hep-th",
    reportNumber = "MZ-TH-08-32",
    doi = "10.1103/PhysRevD.79.025008",
    journal = "Phys. Rev. D",
    volume = "79",
    pages = "025008",
    year = "2009"
}

@article{Manrique:2009tj,
    author = "Manrique, Elisa and Reuter, Martin",
    title = "{Bare versus Effective Fixed Point Action in Asymptotic Safety: The Reconstruction Problem}",
    eprint = "0905.4220",
    archivePrefix = "arXiv",
    primaryClass = "hep-th",
    doi = "10.22323/1.079.0001",
    journal = "PoS",
    volume = "CLAQG08",
    pages = "001",
    year = "2011"
}

@article{Morris:2015oca,
    author = {Morris, Tim R. and Slade, Zo{\"e} H.},
    title = "{Solutions to the reconstruction problem in asymptotic safety}",
    eprint = "1507.08657",
    archivePrefix = "arXiv",
    primaryClass = "hep-th",
    doi = "10.1007/JHEP11(2015)094",
    journal = "JHEP",
    volume = "11",
    pages = "094",
    year = "2015"
}

@article{Fraaije:2022uhg,
    author = "Fraaije, Mathijs and Platania, Alessia and Saueressig, Frank",
    title = "{On the reconstruction problem in quantum gravity}",
    eprint = "2206.10626",
    archivePrefix = "arXiv",
    primaryClass = "hep-th",
    doi = "10.1016/j.physletb.2022.137399",
    journal = "Phys. Lett. B",
    volume = "834",
    pages = "137399",
    year = "2022"
}

@article{Dona:2015tnf,
    author = "Don{\`a}, Pietro and Eichhorn, Astrid and Labus, Peter and Percacci, Roberto",
    title = "{Asymptotic safety in an interacting system of gravity and scalar matter}",
    eprint = "1512.01589",
    archivePrefix = "arXiv",
    primaryClass = "gr-qc",
    doi = "10.1103/PhysRevD.93.129904",
    journal = "Phys. Rev. D",
    volume = "93",
    number = "4",
    pages = "044049",
    year = "2016",
    note = "[Erratum: Phys.Rev.D 93, 129904 (2016)]"
}

@article{Christiansen:2017gtg,
    author         = "Christiansen, Nicolai and Litim, Daniel F. and Pawlowski, Jan M. and Reichert, Manuel",
    title          = "{Asymptotic safety of gravity with matter}",
    eprint         = "1710.04669",
    archivePrefix  = "arXiv",
    primaryClass   = "hep-th",
    reportNumber   = "CERN-TH-2017-216",
    year           = "2018",
    journal        = "Phys. Rev. D",
    volume         = "97",
    number         = "10",
    pages          = "106012",
    doi            = "10.1103/PhysRevD.97.106012"
}

@article{Knorr:2026vax,
    author = "Knorr, Benjamin",
    title = "{Asymptotically (un)safe scattering amplitudes from scratch: a deep dive into the IR jungle}",
    eprint = "2602.21285",
    archivePrefix = "arXiv",
    primaryClass = "hep-th",
    month = "2",
    year = "2026"
}

@inproceedings{Pereira:2019dbn,
    author = "Pereira, Antonio D.",
    title = "{Quantum spacetime and the renormalization group: Progress and visions}",
    booktitle = "{Progress and Visions in Quantum Theory in View of Gravity}: {Bridging foundations of physics and mathematics}",
    eprint = "1904.07042",
    archivePrefix = "arXiv",
    primaryClass = "gr-qc",
    month = "4",
    year = "2019"
}

@article{Pawlowski:2020qer,
    author = "Pawlowski, Jan M. and Reichert, Manuel",
    title = "{Quantum Gravity: A Fluctuating Point of View}",
    eprint = "2007.10353",
    archivePrefix = "arXiv",
    primaryClass = "hep-th",
    doi = "10.3389/fphy.2020.551848",
    journal = "Front. in Phys.",
    volume = "8",
    pages = "551848",
    year = "2021"
}

@inbook{Morris:2022btf,
    author = "Morris, Tim R. and Stulga, Dalius",
    title = "{The Functional f(R) Approximation}",
    eprint = "2210.11356",
    archivePrefix = "arXiv",
    primaryClass = "hep-th",
    doi = "10.1007/978-981-19-3079-9_19-1",
    year = "2023"
}

@inbook{Saueressig:2023irs,
    author = "Saueressig, Frank",
    title = "{The Functional Renormalization Group in Quantum Gravity}",
    eprint = "2302.14152",
    archivePrefix = "arXiv",
    primaryClass = "hep-th",
    doi = "10.1007/978-981-19-3079-9_16-1",
    year = "2023"
}

@inbook{Pawlowski:2023gym,
    author = "Pawlowski, Jan M. and Reichert, Manuel",
    title = "{Quantum Gravity from Dynamical Metric Fluctuations}",
    eprint = "2309.10785",
    archivePrefix = "arXiv",
    primaryClass = "hep-th",
    doi = "10.1007/978-981-19-3079-9_17-1",
    year = "2024"
}

@article{Ellwanger:1993mw,
    author = "Ellwanger, Ulrich",
    editor = "Geyer, B. and Ilgenfritz, E. -M.",
    title = "{FLow equations for N point functions and bound states}",
    eprint = "hep-ph/9308260",
    archivePrefix = "arXiv",
    reportNumber = "HD-THEP-93-30",
    doi = "10.1007/BF01555911",
    journal = "Z. Phys. C",
    volume = "62",
    pages = "503--510",
    year = "1994"
}

@article{Reuter:1996cp,
    author = "Reuter, M.",
    title = "{Nonperturbative evolution equation for quantum gravity}",
    eprint = "hep-th/9605030",
    archivePrefix = "arXiv",
    reportNumber = "DESY-96-080",
    doi = "10.1103/PhysRevD.57.971",
    journal = "Phys. Rev. D",
    volume = "57",
    pages = "971--985",
    year = "1998"
}

@article{Dupuis:2020fhh,
    author = "Dupuis, N. and Canet, L. and Eichhorn, A. and Metzner, W. and Pawlowski, J. M. and Tissier, M. and Wschebor, N.",
    title = "{The nonperturbative functional renormalization group and its applications}",
    eprint = "2006.04853",
    archivePrefix = "arXiv",
    primaryClass = "cond-mat.stat-mech",
    doi = "10.1016/j.physrep.2021.01.001",
    journal = "Phys. Rept.",
    volume = "910",
    pages = "1--114",
    year = "2021"
}

@article{Bonanno:2020bil,
    author = "Bonanno, Alfio and Eichhorn, Astrid and Gies, Holger and Pawlowski, Jan M. and Percacci, Roberto and Reuter, Martin and Saueressig, Frank and Vacca, Gian Paolo",
    title = "{Critical reflections on asymptotically safe gravity}",
    eprint = "2004.06810",
    archivePrefix = "arXiv",
    primaryClass = "gr-qc",
    doi = "10.3389/fphy.2020.00269",
    journal = "Front. in Phys.",
    volume = "8",
    pages = "269",
    year = "2020"
}

@article{Chiesa:2026tlz,
    author = "Chiesa, Angelo P. and Pawlowski, Jan M. and Reichert, Manuel",
    title = "{Towards Two-to-Two Scattering of Scalars in Asymptotically Safe Quantum Gravity}",
    eprint = "2603.10168",
    archivePrefix = "arXiv",
    primaryClass = "hep-th",
    month = "3",
    year = "2026"
}

@article{Pawlowski:2025etp,
    author = "Pawlowski, Jan M. and Reichert, Manuel and Wessely, Jonas",
    title = "{Self-consistent graviton spectral function in Lorentzian quantum gravity}",
    eprint = "2507.22169",
    archivePrefix = "arXiv",
    primaryClass = "hep-th",
    month = "7",
    year = "2025"
}

@article{Braun:2022mgx,
    author = "Braun, Jens and others",
    title = "{Renormalised spectral flows}",
    eprint = "2206.10232",
    archivePrefix = "arXiv",
    primaryClass = "hep-th",
    doi = "10.21468/SciPostPhysCore.6.3.061",
    journal = "SciPost Phys. Core",
    volume = "6",
    pages = "061",
    year = "2023"
}

@article{Eichhorn:2025sux,
    author = "Eichhorn, Astrid and Gyftopoulos, Zois and Held, Aaron",
    title = "{Quark and lepton mixing in the asymptotically safe Standard Model}",
    eprint = "2507.18304",
    archivePrefix = "arXiv",
    primaryClass = "hep-ph",
    month = "7",
    year = "2025"
}

@article{Shaposhnikov:2009pv,
    author = "Shaposhnikov, Mikhail and Wetterich, Christof",
    title = "{Asymptotic safety of gravity and the Higgs boson mass}",
    eprint = "0912.0208",
    archivePrefix = "arXiv",
    primaryClass = "hep-th",
    doi = "10.1016/j.physletb.2009.12.022",
    journal = "Phys. Lett. B",
    volume = "683",
    pages = "196--200",
    year = "2010"
}

@book{Percacci:2017fkn,
    author = "Percacci, Robert",
    title = "{An Introduction to Covariant Quantum Gravity and Asymptotic Safety}",
    doi = "10.1142/10369",
    isbn = "978-981-320-717-2, 978-981-320-719-6",
    publisher = "World Scientific",
    series = "100 Years of General Relativity",
    volume = "3",
    year = "2017"
}

@article{Eichhorn:2012va,
    author = "Eichhorn, Astrid",
    title = "{Quantum-gravity-induced matter self-interactions in the asymptotic-safety scenario}",
    eprint = "1204.0965",
    archivePrefix = "arXiv",
    primaryClass = "gr-qc",
    doi = "10.1103/PhysRevD.86.105021",
    journal = "Phys. Rev. D",
    volume = "86",
    pages = "105021",
    year = "2012"
}

@article{deBrito:2021pyi,
    author = "de Brito, Gustavo P. and Eichhorn, Astrid and Santos, Rafael Robson Lino dos",
    title = "{The weak-gravity bound and the need for spin in asymptotically safe matter-gravity models}",
    eprint = "2107.03839",
    archivePrefix = "arXiv",
    primaryClass = "gr-qc",
    doi = "10.1007/JHEP11(2021)110",
    journal = "JHEP",
    volume = "11",
    pages = "110",
    year = "2021"
}

@article{Knorr:2022ilz,
    author = "Knorr, Benjamin",
    title = "{Safe essential scalar-tensor theories}",
    eprint = "2204.08564",
    archivePrefix = "arXiv",
    primaryClass = "hep-th",
    reportNumber = "NORDITA 2023-064",
    month = "4",
    year = "2022"
}

@article{deBrito:2023myf,
    author = "de Brito, Gustavo P. and Knorr, Benjamin and Schiffer, Marc",
    title = "{On the weak-gravity bound for a shift-symmetric scalar field}",
    eprint = "2302.10989",
    archivePrefix = "arXiv",
    primaryClass = "hep-th",
    reportNumber = "NORDITA 2023-005",
    doi = "10.1103/PhysRevD.108.026004",
    journal = "Phys. Rev. D",
    volume = "108",
    number = "2",
    pages = "026004",
    year = "2023"
}

@article{Eichhorn:2017ylw,
    author = "Eichhorn, Astrid and Held, Aaron",
    title = "{Top mass from asymptotic safety}",
    eprint = "1707.01107",
    archivePrefix = "arXiv",
    primaryClass = "hep-th",
    doi = "10.1016/j.physletb.2017.12.040",
    journal = "Phys. Lett. B",
    volume = "777",
    pages = "217--221",
    year = "2018"
}

@article{Reuter:2004nv,
    author = "Reuter, M. and Weyer, H.",
    title = "{Running Newton constant, improved gravitational actions, and galaxy rotation curves}",
    eprint = "hep-th/0410117",
    archivePrefix = "arXiv",
    reportNumber = "MZ-TH-04-14",
    doi = "10.1103/PhysRevD.70.124028",
    journal = "Phys. Rev. D",
    volume = "70",
    pages = "124028",
    year = "2004"
}

@article{Knorr:2021niv,
    author = "Knorr, Benjamin and Schiffer, Marc",
    title = "{Non-Perturbative Propagators in Quantum Gravity}",
    eprint = "2105.04566",
    archivePrefix = "arXiv",
    primaryClass = "hep-th",
    doi = "10.3390/universe7070216",
    journal = "Universe",
    volume = "7",
    number = "7",
    pages = "216",
    year = "2021"
}

@article{Christiansen:2015rva,
    author = "Christiansen, Nicolai and Knorr, Benjamin and Meibohm, Jan and Pawlowski, Jan M. and Reichert, Manuel",
    title = "{Local Quantum Gravity}",
    eprint = "1506.07016",
    archivePrefix = "arXiv",
    primaryClass = "hep-th",
    doi = "10.1103/PhysRevD.92.121501",
    journal = "Phys. Rev. D",
    volume = "92",
    number = "12",
    pages = "121501",
    year = "2015"
}

@article{Becker:2017tcx,
    author = "Becker, Daniel and Ripken, Chris and Saueressig, Frank",
    title = "{On avoiding Ostrogradski instabilities within Asymptotic Safety}",
    eprint = "1709.09098",
    archivePrefix = "arXiv",
    primaryClass = "hep-th",
    doi = "10.1007/JHEP12(2017)121",
    journal = "JHEP",
    volume = "12",
    pages = "121",
    year = "2017"
}

@article{DAngelo:2025yoy,
    author = {D'Angelo, Edoardo and Ferrero, Renata and Fr{\"o}b, Markus B.},
    title = "{De Sitter quantum gravity within the covariant Lorentzian approach to asymptotic safety}",
    eprint = "2502.05135",
    archivePrefix = "arXiv",
    primaryClass = "hep-th",
    doi = "10.1088/1361-6382/ade193",
    journal = "Class. Quant. Grav.",
    volume = "42",
    number = "12",
    pages = "125008",
    year = "2025"
}

@article{Fehre:2021eob,
    author = "Fehre, Jannik and Litim, Daniel F. and Pawlowski, Jan M. and Reichert, Manuel",
    title = "{Lorentzian Quantum Gravity and the Graviton Spectral Function}",
    eprint = "2111.13232",
    archivePrefix = "arXiv",
    primaryClass = "hep-th",
    doi = "10.1103/PhysRevLett.130.081501",
    journal = "Phys. Rev. Lett.",
    volume = "130",
    number = "8",
    pages = "081501",
    year = "2023"
}

@article{Sen:2022xlp,
    author = "Sen, Saswato and Wetterich, Christof and Yamada, Masatoshi",
    title = "{Scaling solutions for asymptotically free quantum gravity}",
    eprint = "2211.05508",
    archivePrefix = "arXiv",
    primaryClass = "hep-th",
    reportNumber = "YITP-22-130",
    doi = "10.1007/JHEP02(2023)054",
    journal = "JHEP",
    volume = "02",
    pages = "054",
    year = "2023"
}

@article{deAlwis:2017ysy,
    author = "de Alwis, S. P.",
    title = "{Exact RG Flow Equations and Quantum Gravity}",
    eprint = "1707.09298",
    archivePrefix = "arXiv",
    primaryClass = "hep-th",
    doi = "10.1007/JHEP03(2018)118",
    journal = "JHEP",
    volume = "03",
    pages = "118",
    year = "2018"
}

@inproceedings{Glaviano:2024hie,
    author = "Glaviano, E. M. and Bonanno, A.",
    title = "{Proper-time flow equation and non-local truncations in quantum gravity}",
    booktitle = "{17th Marcel Grossmann Meeting}: {On Recent Developments in Theoretical and Experimental General Relativity, Gravitation, and Relativistic Field Theories}",
    eprint = "2410.23696",
    archivePrefix = "arXiv",
    primaryClass = "gr-qc",
    month = "10",
    year = "2024"
}

@article{Bonanno:2019ukb,
    author = "Bonanno, Alfio and Lippoldt, Stefan and Percacci, Roberto and Vacca, Gian Paolo",
    title = "{On Exact Proper Time Wilsonian RG Flows}",
    eprint = "1912.08135",
    archivePrefix = "arXiv",
    primaryClass = "hep-th",
    doi = "10.1140/epjc/s10052-020-7798-9",
    journal = "Eur. Phys. J. C",
    volume = "80",
    number = "3",
    pages = "249",
    year = "2020"
}

@article{Bonanno:2004sy,
    author = "Bonanno, A. and Reuter, M.",
    title = "{Proper time flow equation for gravity}",
    eprint = "hep-th/0410191",
    archivePrefix = "arXiv",
    reportNumber = "MZ-TH-04-16",
    doi = "10.1088/1126-6708/2005/02/035",
    journal = "JHEP",
    volume = "02",
    pages = "035",
    year = "2005"
}

@article{Bonanno:2025tfj,
    author = "Bonanno, Alfio and Oglialoro, Giovanni and Zappal{\`a}, Dario",
    title = "{Gauge and parametrization dependence of quantum Einstein gravity within the proper time flow}",
    eprint = "2504.07877",
    archivePrefix = "arXiv",
    primaryClass = "hep-th",
    doi = "10.1103/sht5-sf25",
    journal = "Phys. Rev. D",
    volume = "112",
    number = "2",
    pages = "026002",
    year = "2025"
}

@article{Codello:2012kq,
    author = "Codello, Alessandro and Zanusso, Omar",
    title = "{On the non-local heat kernel expansion}",
    eprint = "1203.2034",
    archivePrefix = "arXiv",
    primaryClass = "math-ph",
    reportNumber = "MZ-TH-12-12",
    doi = "10.1063/1.4776234",
    journal = "J. Math. Phys.",
    volume = "54",
    pages = "013513",
    year = "2013"
}

@article{Knorr:2019atm,
    author = "Knorr, Benjamin and Ripken, Chris and Saueressig, Frank",
    title = "{Form Factors in Asymptotic Safety: conceptual ideas and computational toolbox}",
    eprint = "1907.02903",
    archivePrefix = "arXiv",
    primaryClass = "hep-th",
    doi = "10.1088/1361-6382/ab4a53",
    journal = "Class. Quant. Grav.",
    volume = "36",
    number = "23",
    pages = "234001",
    year = "2019"
}

@inbook{Knorr:2022dsx,
    author = "Knorr, Benjamin and Ripken, Chris and Saueressig, Frank",
    title = "{Form Factors in Asymptotically Safe Quantum Gravity}",
    eprint = "2210.16072",
    archivePrefix = "arXiv",
    primaryClass = "hep-th",
    reportNumber = "NORDITA 2022-075",
    doi = "10.1007/978-981-19-3079-9_21-1",
    year = "2024"
}

@book{article,
author = {Boyd, John and Marilyn, To and Eliot, Paraphrasing},
year = {2000},
month = {10},
pages = {},
publisher = "Springer Berlin, Heidelberg",
title = {Chebyshev and Fourier Spectral Methods}
}

@article{Bonanno:2021squ,
    author = "Bonanno, Alfio and Denz, Tobias and Pawlowski, Jan M. and Reichert, Manuel",
    title = "{Reconstructing the graviton}",
    eprint = "2102.02217",
    archivePrefix = "arXiv",
    primaryClass = "hep-th",
    doi = "10.21468/SciPostPhys.12.1.001",
    journal = "SciPost Phys.",
    volume = "12",
    number = "1",
    pages = "001",
    year = "2022"
}

@article{Dona:2013qba,
    author = "Don{\`a}, Pietro and Eichhorn, Astrid and Percacci, Roberto",
    title = "{Matter matters in asymptotically safe quantum gravity}",
    eprint = "1311.2898",
    archivePrefix = "arXiv",
    primaryClass = "hep-th",
    doi = "10.1103/PhysRevD.89.084035",
    journal = "Phys. Rev. D",
    volume = "89",
    number = "8",
    pages = "084035",
    year = "2014"
}






\end{document}